\documentstyle[12pt,aaspp4]{article}
\newcommand{\oi}{\ion{O}{1}}
\newcommand{\fev}{\ion{Fe}{5}}
\newcommand{\feiv}{\ion{Fe}{4}}
\newcommand{\feiii}{\ion{Fe}{3}}
\newcommand{\nai}{\ion{Na}{1} }
\newcommand{\ea}{{et~al.}}

\newcommand{\IUE}{{\it IUE}}

\newcommand{\logg}{$\log g$}

\newcommand{\lsun}{L$_{\sun}$}
\newcommand{\msun}{M$_{\sun}$}
\newcommand{\mv}{{M}$_{V}$}
\newcommand{\mbi}{m$_{152}$}
\newcommand{\mbv}{m$_{162}$}

\newcommand{\teff}{T$_{eff}$}

\begin{document}  

\title{The Hot Stars of Old Open Clusters: M67, NGC 188 and NGC~6791}

\author{Wayne Landsman\altaffilmark{1},  Ralph C. Bohlin\altaffilmark{2}, Susan 
G. Neff \altaffilmark{3},
Robert W. O'Connell\altaffilmark{4}, Morton S. Roberts\altaffilmark{5},
Andrew M. Smith\altaffilmark{3}, and Theodore P. Stecher\altaffilmark{3}}

\altaffiltext{1}{Raytheon STX Corporation, NASA Goddard Space Flight Center,
Laboratory for Astronomy and Solar Physics, Code 681, Greenbelt, MD 20771}

\altaffiltext{2}{Space Telescope Science Institute, 3700 San Martin Drive,
Baltimore, MD 21218}

\altaffiltext{3}{NASA Goddard Space Flight Center,
Laboratory for Astronomy and Solar Physics, Code 681, Greenbelt, MD 20771}

\altaffiltext{4}{University of Virginia, Department of Astronomy, 
P.O. Box 3818, Charlottesville, VA 22903}

\altaffiltext{5}{National Radio Astronomy Observatory, 520 Edgemont Rd.,
Charlottesville, VA 22903-2475}
\vspace{0.1in}

\begin{abstract}

   We analyze ultraviolet ($\sim 1500$~\AA) images of the old open clusters
M67, NGC 188, and NGC 6791 obtained with Ultraviolet Imaging Telescope (UIT)
during the second flight of the {\em Astro} observatory in March 1995.    
Twenty stars are detected on the UIT image of M67, including 11 blue
stragglers, seven white dwarf candidates, and the yellow giant -- white dwarf
binary S1040.     
The
ultraviolet photometry of the blue stragglers F90 (S975) and F131 (S1082)
suggests that these stars have hot subluminous companions.
We present a semi-empirical integrated ultraviolet
spectrum of M67, and show that the blue stragglers  
dominate the integrated spectrum at wavelengths shorter than 2600~\AA.
The number of white
dwarfs in M67 is roughly consistent with the number expected from white dwarf
cooling models. Eight candidate sdB/sdO stars are detected in NGC 6791, and two
are detected in NGC 188.    The luminosity range $1.10 < \log$ L/\lsun\ $<
1.27$, derived from the ultraviolet photometry of  the six sdB candidates, is
consistent with theoretical models of metal-rich hot horizontal branch (HB)
stars.  The fraction of hot HB stars in both NGC 6791 and NGC 188 is about
30\%, implying that the integrated spectra of both clusters should show a UV
turnup at least as strong as that  observed in any elliptical galaxy.

\end{abstract}

\keywords{open clusters and associations: individual (M67, NGC 188, NGC 6791)
 -- stars: white dwarfs --- stars: blue stragglers --- ultraviolet: stars}

\section{Introduction} As a stellar population ages, the main-sequence turnoff
becomes cooler and fainter, until, after an age of about 2 Gyrs, the
ultraviolet  ($\sim$ 1500~\AA) flux is no longer dominated by stars near the
main-sequence, but instead must be due to a more exotic  population of hot
stars, such as blue stragglers, hot subdwarf (sdB, sdO) stars, hot
post-asymptotic giant branch (post-AGB) stars, white dwarfs, or interacting
binaries.    The blue stragglers and hot subdwarfs are of particular interest
since they might have significant roles, respectively, in two current problems
in extragalactic astronomy:  the age-dating of old galaxies from their
rest-frame near-UV ($\sim 2600$~\AA) spectra,  and understanding the origin of
the observed UV-upturn in elliptical galaxies.      In the absence of blue
stragglers, the rest-frame near-UV spectrum of an old galaxy should be
dominated by the turnoff population, and thus fitting the spectrum to
population synthesis models should be a reliable method for age-dating the
galaxy, although metallicity effects also need to be properly disentangled
(\cite{heap98}, \cite{bruz97}).       However, as noted by  
\markcite{spin97}Spinrad et al. (1997), a significant underestimate of the
galaxy age could result if blue stragglers  contribute a large fraction of the
galaxy near-UV flux, but are not included in the adopted population synthesis
model.        

\markcite{greg90}Greggio \& Renzini (1990) outlined the various candidates for
the origin of the observed UV-upturn in elliptical galaxies, which is
characterized by a rising UV flux shortward of 1800~\AA\ (\cite{bica96b},
\cite{brown97}).   Among these candidates, white dwarfs and hot post-AGB stars
have the advantage of being relatively well-understood phases of stellar
evolution, but lack sufficient fuel to explain the strongest UV-upturn
galaxies.   The more promising candidates are currently the extreme
horizontal-branch (EHB) stars and their hot progeny; the AGB-manqu\'e stars,
which miss the AGB phase entirely, and the post-early AGB stars, which leave
the AGB before the thermally pulsing phase.     Although the horizontal branch
(HB) morphology of the globular clusters becomes generally redder with
increasing metallicity, a significant EHB population is theoretically expected
at metallicities above solar (and an age larger than about 5 Gyr), if reasonable
assumptions are made concerning the metallicity dependence of the mass loss
rate and helium abundance  (\cite{dorm95}, \cite{yi97a}). Spectroscopically,
EHB and AGB-manqu\'e stars are expected to appear, respectively, as  sdB and
sdO stars, and the measured effective temperatures and gravities of the  field
hot subdwarf stars is consistent with this scenario (\cite{saff94}).   However,
the field hot subdwarfs apparently have a large binary fraction
(\cite{theis95}, \cite{green98b}), which instead suggests that binary
interactions might be involved in their formation (\cite{bailyn95}).

Despite their probable contributions to the ultraviolet spectra of galaxies,
blue stragglers and hot subdwarfs  are either neglected or poorly constrained
in existing population synthesis models of old stellar populations.    The
currently  favored explanations for the origin  of blue stragglers involve
binary interactions, either through mass-transfer processes,  binary mergers,
or direct stellar collisions  (\cite{bailyn95}, \cite{leonard96}), but the
large parameter space of these binary  processes limits their inclusion in
synthesis models to special cases (e.g.\ \cite{pols94}).    Some attempts have been
made to include EHB stars in synthesis models of metal-rich populations, but
these are  hampered by  large uncertainties, for example, in the dependence of 
mass loss and helium abundance on metallicity (\cite{bress94}, \cite{yi97b}). 
The modeling of galaxy ultraviolet spectra should clearly benefit from empirical
studies of the ultraviolet content of resolved, old,  metal-rich stellar 
populations.    

Here we report on ultraviolet ($\sim$ 1600~\AA) images of the three old ($> 4$
Gyr) open  clusters M67, NGC 188, and NGC 6791, which were obtained in March
1995  using the {\em Ultraviolet Imaging Telescope} (UIT, \cite{stech97})
during the second flight of the {\em Astro} observatory.      The $40'$
diameter field
of  view of UIT is sufficiently large to allow a complete  census of the hot
stellar population in the observed clusters.   Thus, the UIT images of these
three clusters can be used for an  empirical study of the ultraviolet content
of old, metal-rich stellar populations.     As it turns out, M67 contains blue
stragglers, but no hot subdwarfs, whereas NGC 188 and NGC 6791 contain hot
subdwarfs, but  are too old, distant and reddened for the blue stragglers to be
detectable with UIT.    M67 is also sufficiently nearby that all the white
dwarfs hotter than $\sim$ 21,000 K should have been detected on the UIT
image.   In the population synthesis models of \markcite{magris93}Magris \&
Bruzual (1993),  white dwarfs supply about 10\% of the 1750~\AA\ flux in
elliptical galaxies with a weak UV-upturn,  and in Section 3.2 we re-examine  
the white dwarf contribution to galaxy ultraviolet spectra, using more recent
stellar atmospheres and evolutionary models.

A brief summary of this work was given by \markcite{last97}Landsman \&
Stecher (1997). A subsequent paper will discuss the UIT observations of the
intermediate age ($\sim$ 2 Gyr) clusters NGC 752 and NGC 7789, which are
sufficiently young to allow the turnoff population to be detected in the
ultraviolet. 

\section{Observations}

\begin{figure}
\epsscale{0.80} 
\plotone{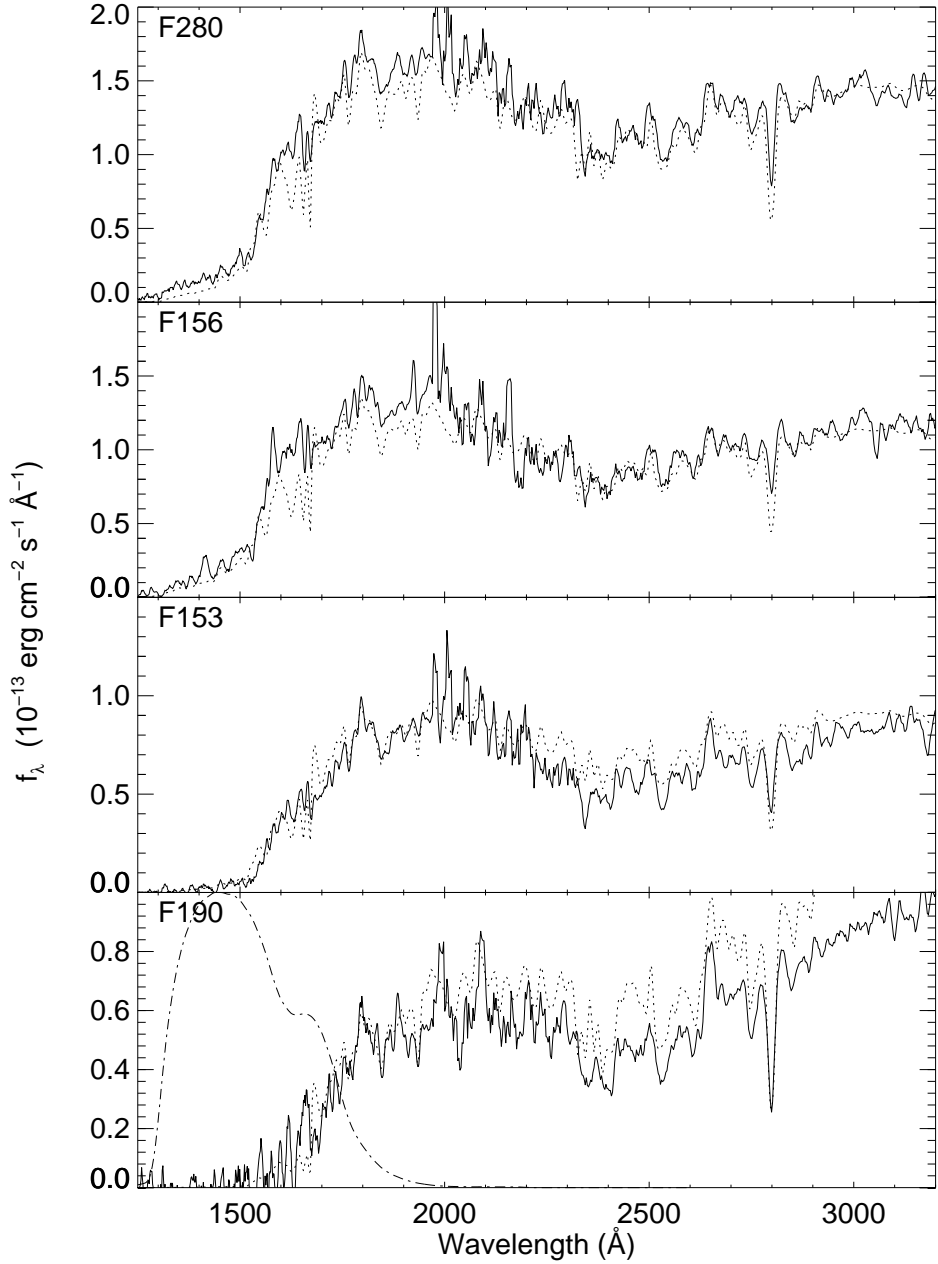} 
\figcaption{IUE spectra of the blue stragglers
F280, F156, F153, and F190 (solid lines) are compared with Kurucz model spectra
(dotted lines), normalized to the V magnitudes, reddened by E(\bv) = 0.025,
and with [Fe/H] = --0.1, and \teff\ and \logg\ derived from Str\"omgren
photometry.    No adjustment to the parameters were made to improve the fit.
The filter response of the UIT B1 filter is indicated by a dot-dashed line on
the spectrum of F190. \label{fig-3}}  

\end{figure}

Table \ref{tbl-1} gives the adopted fundamental parameters of the observed clusters, and
Table \ref{tbl-2}  lists the deepest exposure obtained with each UIT filter, along with
the coordinates of the UIT image center.    The UIT image center was chosen to
provide a target for the co-pointed spectroscopic instruments on the {\em
Astro} observatory, and could be offset by as much as $11'$ (in the case of 
NGC 6791)
from the cluster center. M67 was observed with the B1 filter, which has a peak
wavelength of 1520~\AA\ and a bandwith of 350~\AA\ (see Figure \ref{fig-3}).   NGC 6791
was observed with the B5 filter, which has a  peak wavelength of 1620~\AA\ and
a bandwidth of  230~\AA.    NGC 188 was observed with both the B1 and the B5
filter.     (The less sensitive B5 filter was used on the daytime side of the
Shuttle orbit, because it suppresses \ion{O}{1} 1300~\AA\ dayglow emission.) 
Ultraviolet magnitudes on the B1 and B5 images are denoted here as \mbi\ and 
\mbv, respectively, and are given on the monochromatic system,
where $m_{\lambda} = -2.5 \log(F_{\lambda}) -21.1$, and $F_{\lambda}$ is the
observed flux in units of erg cm$^{-1}$ s$^{-1}$ \AA$^{-1}$. 

Only a few stars in each cluster are detected on the UIT images, so that
ultraviolet magnitudes could be derived from simple circular aperture
photometry.    However, the absolute calibration of the UIT images is somewhat
problematic.    As discussed by \markcite{stech97}Stecher et al.\ (1997),
comparison of fluxes from UIT and  the {\em International Ultraviolet Explorer}
(IUE)  indicated that the sensitivity of the UIT during the {\em Astro-2}
mission appeared to decrease with exposure time.   The cause of this behaviour
has not been determined, and instead an empirical correction to the exposure
time was adopted that, for most images, brought the IUE and UIT fluxes to
within 15\% of each other.    For M67, this mean calibration could be tested,
and an aperture correction determined, by direct comparison with spectra from
IUE and the {\em Goddard High-Resolution Spectrograph} (GHRS)  of five stars in
the cluster.    The NGC 6791 calibration was modified by comparison with a
single star (NGC 6791-B4) with a GHRS observation, while for NGC 188 we had to
rely on the mean UIT calibration.   Thus while the photometry of all the
clusters has a random uncertainty of about 0.1 mag, we estimate an uncertainty
in the absolute photometry  of about 0.1 mag for M67, and about 0.15 mag for
NGC 6791 and NGC 188.   

All IUE spectra in this paper used the NEWSIPS (\cite{nichol96})
reduction, with the fluxes multipled by 1.06, as suggested by 
\markcite{col94}Colina \& Bohlin (1994).

\section{M67}

\begin{figure}  
\plotone{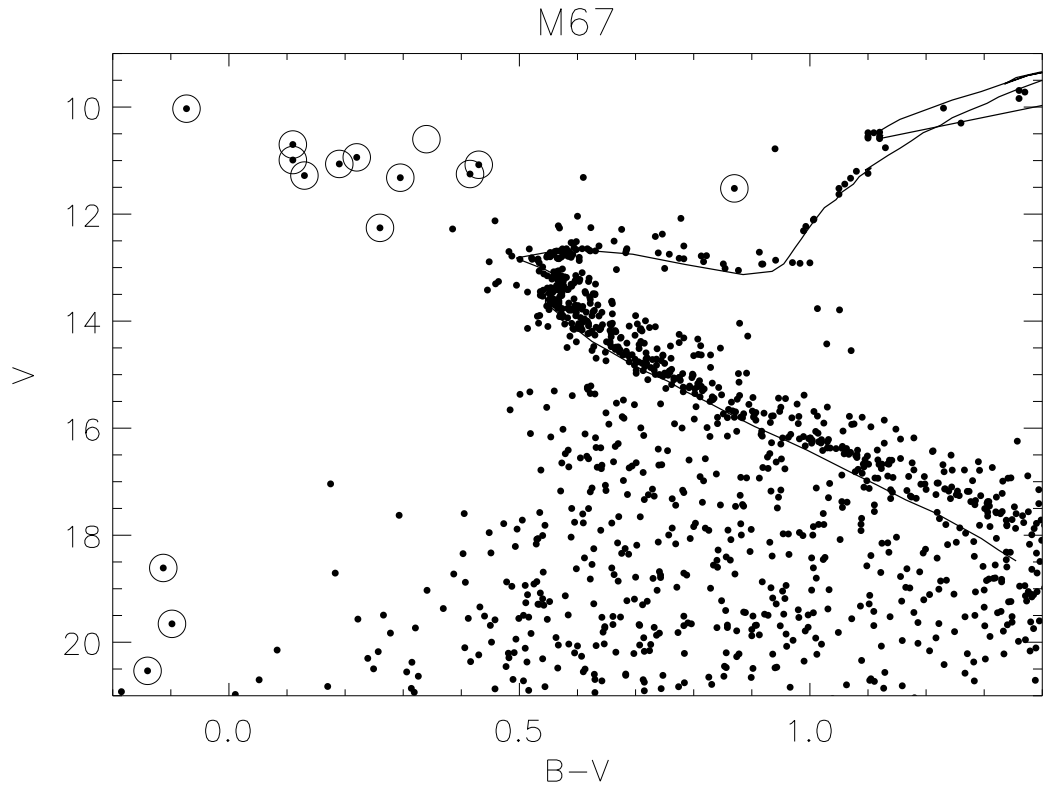}   
\figcaption{ The visible color-magnitude
diagram of M67 from MMJ93.     The bright non-members (mainly stars with V $<
15$) from the proper motion study of Girard \ea\ (1989) have been removed.  
Stars detected on the 1520~\AA\ UIT image are circled.    The probable
non-member S1466 is shown in the blue straggler region as a circle without a
central dot.   The solid line shows a 4 Gyr solar-metallicity isochrone from
Bertelli et al.\ (1994) with m--M = 9.65 and E(\bv) = 0.025. \label{fig-2}} 
\end{figure}

M67 is a well-studied, nearby (820 pc), solar-metallicity open cluster with an
age of about four Gyr.  Twenty stars are detected on the UIT image of M67 (Figure
\ref{fig-1}), including  eleven blue stragglers, seven white dwarf candidates,
the yellow-giant S1040 (from the catalog of \cite{sand77}), and the probable
non-member S1466.      The
integrated 1520~\AA\ flux of M67 is completely dominated by the blue
stragglers, and in particular, by the bright,  hot (V = 10.03, \bv = --0.09)
blue straggler F81.   Figure \ref{fig-2} shows a color-magnitude diagram (CMD)
of a $30'$ field centered on M67 taken from \markcite{mont93}Montgomery et al.\
(1993, hereafter MMJ93), with circles identifying the sources detected in the
ultraviolet.   In general, the stars that are detected in the ultraviolet  are
ones with the bluest \bv colors, with the notable exception of the star S1040,
which has V = 11.52 and  \bv = 0.82.    S1040 is a known spectroscopic binary
and \markcite{land97}Landsman et al.\ (1997) used GHRS spectroscopy of S1040 to
show that the ultraviolet flux is due to a hot white dwarf companion of the
yellow giant.  Landsman et al.\ also constructed a mass-transfer history for
S1040 that explains its unusual ``red straggler'' location in the CMD (0.2 mag
blueward of the giant branch), relatively long orbital period (42d), and the
unusually low mass ($\sim 0.23$ \msun) of the white dwarf companion.   In their
scenario, S1040 originated as a short-period ($\sim$ 2d) binary, in which the
more massive star (the donor) fills its Roche lobe while on the lower giant
branch.  A period of rapid mass transfer  reverses the mass-ratio of the donor
and accretor (but without creating a common envelope), and is followed by an
extended period ($\sim$ 800 Myr) of stable mass transfer.  During the mass
transfer, the orbital period increases, and the accretor eventually becomes a
blue straggler, more massive than the cluster turn-off stars.   Mass transfer
ends when the donor's envelope is depleted, leaving a low mass helium white
dwarf, and a blue straggler, which has since evolved redward to its current
location in the CMD.

\subsection{Blue Stragglers}

Eleven of the thirteen blue stragglers in M67 listed by
\markcite{\mila94}Milone \& Latham (1994) are detected on the UIT image.  The
first two columns of Table \ref{tbl-3} contain, respectively, the  Fagerholm
and  the \markcite{sand77}Sanders (1977) identifications of the detected blue
stragglers, the third and fourth columns list the V and
\bv\ magnitudes tabulated by MMJ93, and the fifth column gives the observed UIT
magnitude, \mbi.   The brightest and hottest blue straggler is F81 (V = 10.03,
\bv = --0.07), which is saturated even on the shortest exposure (109s) UIT
image.     Optical and IUE spectra of F81 were studied by
\markcite{schon94}Sch\"onberner \& Napiwotzki (1994), who derived \teff\ =
12,750 K, \logg\ = 4.26 and a spectroscopic mass of $3.1 \pm 0.3$ \msun.  
Comparing the IUE spectra of F81 with the UIT photometry of the remaining stars
in M67, we find that F81 supplies more than 90\% of the total 1520~\AA\ flux of
M67.    Despite the extreme properties of F81, the evidence is strong from both
astrometric (\cite{girard89}) and radial velocity (\cite{milone92}) studies
that it is a cluster member.

In order to search for evidence of a subluminous hot companion, or an unusual
ultraviolet spectrum, we compare the UIT photometry of the blue stragglers with
model atmosphere predictions, using stellar parameters derived from optical
photometry.   Temperatures and gravities for most of the blue stragglers are
calculated from the Str\"omgren photometry tabulated by 
\markcite{hauck98}Hauck \& Mermilliod (1998), using the code of 
\markcite{nap93}Napiwotzki et al.\ (1993), and assuming E(b--y) = 0.73E(\bv) =
0.018 for all stars.   No Str\"omgren photometry is available for F90,  and so
the values of \teff\ and \logg\ for this star are taken from
\markcite{mathys91}Mathys (1991), who derived them using Geneva photometry.   
The  predicted \mbi\  magnitude is then calculated from the derived \teff\ and
\logg\ using the 1995 Kurucz LTE model atmospheres as tabulated by  Lejeune et
al.\ (1997).  This version of the Kurucz models includes the modification of 
the convective overshoot algorithm discussed by \markcite{cast96}Castelli
(1996),  which eliminates the discontinuities in color indicies for 6500~K $<$
\teff $<$ 8000~K that was present in earlier versions of the Kurucz grid.  The
models are  normalized to the stellar V magnitude, and reddened using the
parameterization of  \markcite{card89}Cardelli et al.\ (1989) with R$_{V}$ =
3.1.   The calculated values of \teff\ and \logg\ are listed in columns 5 and 6
of Table \ref{tbl-3}, and the predicted \mbi\ magnitudes are given in column 7.

For the range of effective temperatures considered here (6500 -- 8500 K), the
UIT 1520~\AA\ bandpass is located shortward of the blackbody peak, and thus is
very sensitive both to the exact value of \teff\, and to the accuracy of the model
atmosphere.    An uncertainty of 200 K in the value of \teff\ corresponds to
about a 0.5 mag uncertainty in the predicted UIT magnitude. 
Archival IUE spectra are available for the blue stragglers F280, F156, F153,
and F190, and these provide a much more robust comparison with the model
atmospheres.    Figure \ref{fig-3} compares the IUE spectra with Kurucz models with [Fe/H] =
--0.1, and \teff\ and \logg\ from Table \ref{tbl-3}, normalized to the stellar V
magnitude.   No parameters were adjusted to improve the fit.  The model spectra
provide a good fit to both F280 and F156, and the model fits for F153 and
F190 can be improved by adopting values of \teff\ that are lower by 60 K and
140 K, respectively, than the values listed in Table \ref{tbl-3}; a change that is within
the uncertainty of the Str\"omgren photometry.   In addition, the somewhat
poorer fit for F153 may be related to its Am star abundance peculiarities
(\cite{mathys91}),  while the poorer fit for F190 may be related to its 0.02
mag $\delta$ Sct  pulsations (\cite{gibr92}), since the flux variations of
$\delta$ Sct stars are known to be much larger in the ultraviolet
(\cite{mon94}).   We conclude that the IUE spectra provide no evidence for a
subluminous companion in any of the four stars.  The lack of a hot subluminous
companion for F190 is especially noteworthy, because it is a  single-lined
spectroscopic binary with a 4.2d period, and suspected to be currently
undergoing mass transfer (\cite{mila92}).   We also note that if we let the
reddening be a free parameter in the model atmosphere fits, then lower
reddening values (0.015 $<$ E(\bv) $<$ 0.025) are consistently favored over the
higher values in the range usually considered acceptable for M67 (0.015 $<$
E(\bv) $<$ 0.052; \cite{fan96}).   

For the remaining blue stragglers, taking into account the $\sim 0.5$ mag
uncertainty in the predicted UIT flux, and the 0.14 mag uncertainty in the
measured UIT flux, there is strong evidence for a UV excess only for the two
stars F131 (S1082) and F90 (S975). In the case of F131, the observed 1520~\AA\
flux is a factor of six higher than predicted from the optically derived
\teff.    \markcite{mathys91}Mathys (1991) found evidence for an additional
broad component in the \nai\ D and \oi\ absorption lines of F131,  which he
suggested was due to a hot, fast-rotating secondary.  F131 is also a ROSAT
source with an X-ray luminosity of $4 \times 10^{30}$ erg s$^{-1}$
(\cite{bell93}).     The X-ray emission and the presence of a hot, subluminous
secondary together suggest that  F131 is an Algol-type mass-transfer binary.   
Milone (1991) reports that F131 is a radial velocity variable, with a small
amplitude and a long period, but no orbit has yet been determined.      

The UIT magnitude of F90 is probably uncertain  by close to a factor of two,
because F90 is located in the wings of the heavily saturated image of F81.
Nevertheless, the detection of F90 with UIT is quite robust, and likely due to
a hot subluminous companion, since the predicted ultraviolet magnitude from the
optically-derived \teff\ is about 1.9 mag fainter than the UIT detection
limit.  \markcite{lath96}Latham \& Milone (1996) report that S975 is a binary
with a period of 1221 days, and a small eccentricity $e = 0.088 \pm 0.060$.    
The probable presence of a hot subluminous companion in F90, along with a
nearly circular binary orbit, suggests that its blue straggler origin is also
due to a binary mass-transfer process.   Note that while the hot companion of
F131 is probably not a white dwarf (since it is detected in the  \nai\ D and
\oi\ absorption lines), the nature of the hot companion of F90 is unknown.  

The star S1466 occupies the blue straggler region of the HR diagram, and is
located $7'$ from the center of M67.     However, the star is usually considered
to be a non-member based on proper motion studies.  The most precise proper
motion study of M67 is that of 
\markcite{girard89}Girard et al.\ (1989), who give a 21\%
probability of cluster membership for S1466.     Further study of S1466 is
perhaps warranted, given that a blue straggler might acquire a peculiar velocity
if its formation process is due to stellar interactions 
(e.g.\ \cite{leonard96}).

\begin{figure} 
\epsscale{0.80}
\plotone{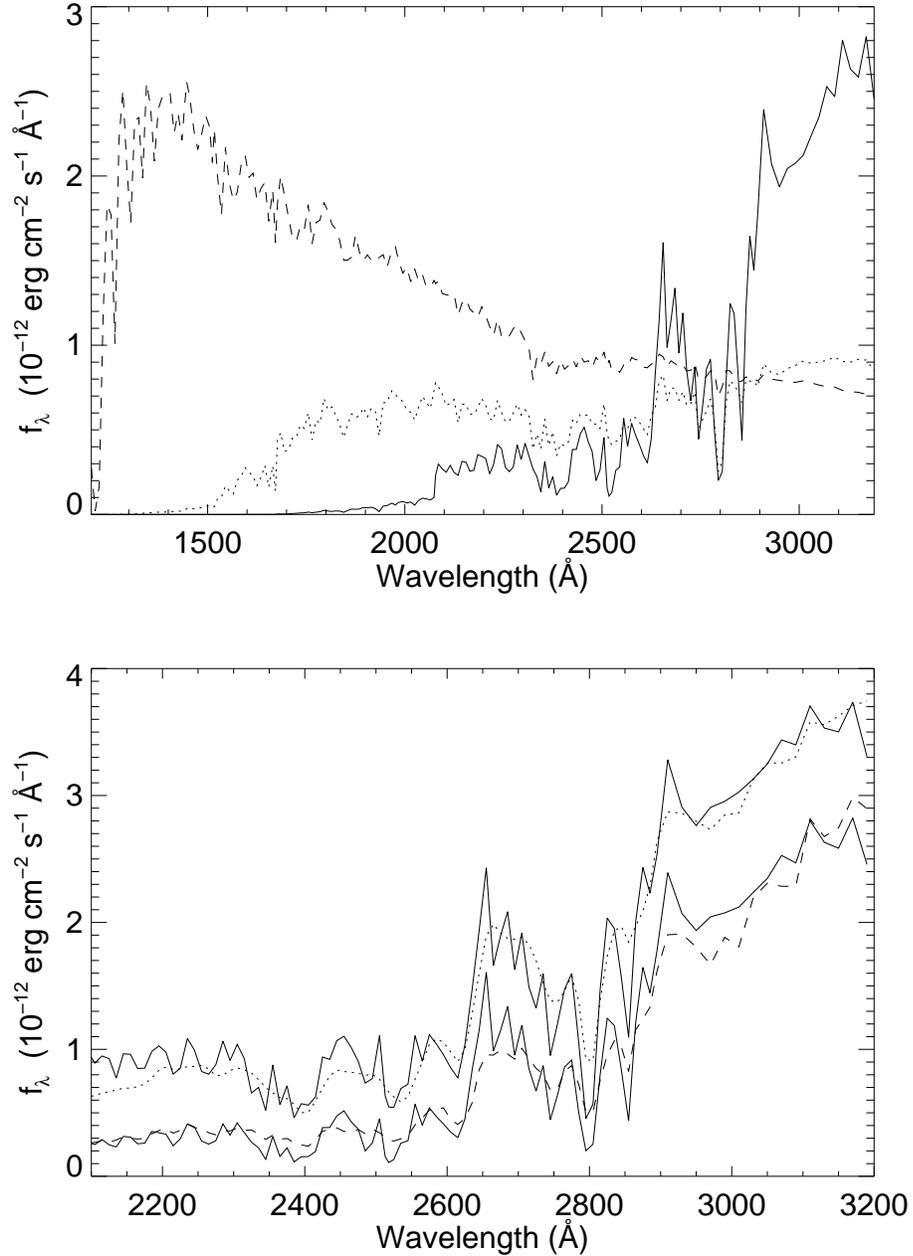} 
\figcaption{The integrated spectrum of M67
computed with Kurucz model atmospheres, using temperature and luminosities
derived from the V and  \bv colors of the proper-motion members.   In the upper
panel, the integrated M67 spectrum is broken into three components: (1)  dashed
line -- the bright blue straggler F81, (2) dotted line -- the remaining ten
blue stragglers, and (3) solid line -- all other stars in the cluster.   In the
lower panel, the lower solid line shows the integrated spectrum excluding all
blue stragglers, and the dashed line shows the best-fit Bruzual \& Charlot
synthesis spectrum,
which has an age of 4.5 Gyr.   The upper solid lines show the integrated
spectrum excluding only F81, and the dotted line shows the best-fit Bruzual \&
Charlot synthesis spectrum, which has an age of  2.5 Gyr. 
\label{fig-4}}    
\end{figure}

To study the contribution of blue stragglers at other wavelengths, we perform a
semi-empirical computation of the integrated M67 ultraviolet spectrum, using
the list of stars with at least 80\% probability of cluster membership from the
astrometric studies  of (in order of preference)  \markcite{girard89}Girard et
al. (1989), \markcite{zhao93}Zhao et al.\ (1993), and \markcite{sand77}Sanders
(1977).     Although the list of proper motion members does not extend fainter
than about V = 15 (Figure \ref{fig-2}), the fainter main-sequence stars should contribute
negligible light below 3000~\AA.   For all but two of the proper motion
members, V and \bv\ magnitudes could be obtained from the catalogs of MMJ93,
\markcite{sand89}Sanders (1989), \markcite{girard89}Girard et al.\ (1989) or
\markcite{fan96}Fan et al.\ (1996).   Effective temperatures are estimated from
the dereddened \bv\ colors,  using  the \teff\ -- \bv\ relation  of
\markcite{lej97}Lejeune et al.\ (1997).    Gravities are approximated by
assuming a mass of 1.25 \msun\ for stars brighter than V=13, and assuming
\logg\ = 4.5 for fainter stars.     From the \teff\ and \logg\ of each star,  a
model synthetic ultraviolet spectrum from the  compilation of Lejeune et al.\
(1997) is assigned, normalized to the star's dereddened V magnitude.     
Figure \ref{fig-4} shows the predicted  integrated ultraviolet spectrum of M67, with the
blue straggler  contribution shown separately from all other stars.   Even
neglecting the possibly anomalous star F81,  the  blue straggler contribution
still dominates the integrated M67 spectrum  at wavelengths shorter than
2600~\AA.  

The lower panel of  Figure \ref{fig-4} compares the integrated  spectra with
synthetic spectra derived from the solar-metallicity isochrones computed by
Bruzual \& Charlot, and tabulated in  \markcite{leith96}Leitherer et al.\ (1996,
also see \cite{charl96}).   The synthetic spectra are computed  for an
instantaneous starburst and a Salpeter initial mass function.       (We do not
attempt to fit the integrated spectra shortward of 2100~\AA, because the
Bruzual \& Charlot models include a contribution from the short-lived hot
post-AGB stars, which do not exist  in the relatively small population of M67.)
If the blue stragglers are excluded, than a 4.5 Gyr model provides the
best fit to the semi-empirical integrated spectrum, which is in agreement with
the age  of M67 estimated from color-magnitude diagrams.   However, if all the
blue stragglers except F81 are included in the integrated spectrum, then a 2.5
Gyr isochrone provides the best fit, and, in particular, provides a good fit of
the spectral break at 2600~\AA.    Thus, in this case, the neglect of
blue stragglers in a population synthesis model of the unresolved ultraviolet
spectrum would lead to a significant underestimate of the cluster age.

There are at least two ways in which the blue straggler population of M67 
might differ from that of galaxies of similar age.    First, dynamical
interactions of M67 with the Galaxy have enhanced the relative contribution of
blue stragglers to the integrated spectrum, because the low-mass stars have
evaporated from the cluster, while blue stragglers are concentrated toward the
cluster center (MMJ93).    A more important question concerns the origin of
blue stragglers themselves; whether they arise from primordial binaries either
via mass transfer processes or stellar mergers, or whether they originate from
stellar collisions enhanced by the moderately high stellar density (20
pc$^{-3}$; \cite{leonard96}) in the M67 center.    While this stellar density
is often exceeded in the cores of many elliptical galaxies (\cite{lauer95}),
the typical star density in elliptical galaxies or stellar bulges will be at
least an order of magnitude lower.

In fact, there is good evidence that both mass-transfer and stellar collision
processes contribute to the M67 blue straggler population.     The short-period
(4.2d) binary F190 has long been a good candidate for ongoing mass transfer
(\cite{mila92}), and the present work provides evidence that both F90 and F131
are Algol-type mass-transfer systems.  As noted above, 
\markcite{land97}Landsman et al.\ (1997)
showed that the yellow giant S1040 is almost certainly a blue straggler
descendant, and a product of early case B mass transfer.  On the other hand,
the bright blue straggler F81 has a mass larger than twice the turnoff mass,
and the blue stragglers F55, F238, and F555 (S997) have orbits with appreciable
eccentricities (\cite{lath96}) . Neither of these properties can be easily
understood with a mass-transfer scenario, or mergers of primordial binaries, 
but are easily  accommodated with binary-binary or binary-single star collisions
(\cite{leonard96}).   

Is the large contribution of blue stragglers to the integrated ultraviolet
spectrum of M67 typical of other open clusters? Spinrad et al.\ (1997)
estimate that the blue straggler contribution to the cluster ultraviolet
spectrum is higher for M67 than for any of the seven clusters they examined.
Ahumada \& Lapasset (1995) have compiled a catalog of blue stragglers in 390
open clusters, in which  they tabulate both Nbs, the number of blue stragglers
in the cluster, and N2, the number of stars on the main sequence to two
magnitudes below the turnoff.    The ratio Nbs/N2 is 0.15 for M67, about twice
the mean of the entire open cluster population.   We do not pursue this
comparison any further here, but instead make the following two points.   
First, the best way to detect the presence of  blue stragglers in an unresolved
population is to have wide spectral coverage, including both ultraviolet and
visible bandpasses.    Second, evolutionary models of close binaries including
mass transfer (e.g.\ \cite{pols94}) would be a useful addition to galaxy
population synthesis  models, especially for those spectral regions expected to
be dominated by the stellar turn-off population.  

\subsection{White Dwarfs}

Figure \ref{fig-2} shows that three UIT sources are coincident with faint hot
sources in MMJ93, and occupy the white dwarf region in the lower left-hand
corner of the HR diagram.   Four other UIT sources are outside of the region
studied by MMJ93, but are coincident with faint, hot stars in the BATC 
(Beijing-Arizona-Taipei-Connecticut) intermediate-band spectrophotometric
survey  of \markcite{fan96}Fan et al.\ (1996).      The UIT photometry of these
seven white dwarf candidates, along with that of the yellow giant -- white
dwarf binary S1040 is given in column 4 of Table \ref{tbl-4}.   The BV
photometry is given in columns 2 and 3, where the  intermediate-band photometry
of the BATC survey is converted to V and \bv magnitudes using equations (3) and
(4) in \markcite{fan96}Fan et al.\ (1996).    Column 5 gives the distance of
the white dwarf candidates from the star S1023, assumed to be the cluster
center. 

The brightest
of the white dwarf candidates (and the fifth brightest source in M67 at
1520~\AA) is MMJ 5670 (V $\sim 18.6$), which is also star G152 in 
\markcite{gill91}Gilliland et
al.\ (1991).    This star was suspected to be responsible for  the soft
component of a blended ROSAT X-ray source in M67 discovered by
\markcite{bell93}Belloni et al.\ (1993).    Subsequent  optical spectroscopy by
\markcite{pasq94}Pasquini et al.\ (1994) and \markcite{flem97}Fleming et al.\
(1997) demonstrated that MMJ5670 is a hot DA white dwarf.    Fleming et al.\
derive \teff\ = 68,230 $\pm 3200$ K, and \logg\ = $7.58 \pm 0.16$ from a model 
atmosphere fit to the Balmer lines.    Using a pure hydrogen model atmosphere
with this \teff\ and \logg\, and normalizing to V = 18.71, and assuming E(\bv)
= 0.025, yields a  predicted UIT magnitude of \mbi\ = 14.11, in reasonable
agreement with the observed value of \mbi = 13.92.      Only 10\% of hot ($>$
20,000 K) white dwarfs are X-ray luminous (\cite{flem96}), and this fraction is
much lower for white dwarfs with  \teff\ $>$ 50,000 K, because the X-ray
emission in hot white dwarfs can be strongly suppressed by the presence of
radiatively supported trace metals in the photosphere (\cite{marsh97},
\cite{wolff98}).     Thus, MMJ 5670 must have have a remarkably pure hydrogen
atmosphere.     The high \teff\ of MMJ 5670 also indicates a young age;
according to the 0.6 \msun\ white dwarf cooling models of Wood (1995), MMJ 5670
began its descent on the white dwarf cooling curve within the past $10^6$ yr.

\markcite{flem97}Fleming et al.\  (1997) also obtained an optical spectrum of MMJ 5973 (V $\sim
19.7$),  after learning of its detection on the UIT image.    
They find that MMJ 5973
has a  DB spectrum, and they derive \teff\ = 17,150 $\pm 310$ K, and \logg\ =
$7.77 \pm 0.47$  from a model atmosphere fit.    
Presuming cluster membership
(see below), MMJ 5973 is the first DB white dwarf found in an open cluster, and
shows that both DB and DA white dwarfs can be produced in the same cluster.   
The origin of the distinction between DA and DB white dwarfs is still a matter
of some debate.    The high surface helium abundance of DB  white dwarfs
might arise if the DB stars are descended from stars that left the  asymptotic
giant branch (AGB) during a helium shell flash (e.g.\ \cite{deka95}).   As
opposed to this ``primordial'' scenario,  
surface phenomena such as mass loss,  gravitational
settling and convective overturn might provide an evolutionary path between DB 
and DA white dwarfs, provided that the hydrogen surface layer is
sufficiently thin (\cite{fowe97}).    

Remarkably, all three of the spectroscopically studied white dwarfs in M67 
have turned out to have unusual characteristics; MMJ 5670 is a very hot 
X-ray-luminous DA, MMJ 5973 is a DB, and the S1040 companion is a  very low 
mass core-helium white dwarf.      No spectra exist for the five faintest white
dwarf candidates in Table \ref{tbl-4}.    The three faintest sources (MMJ 6061,
BATC 2776, BATC 3337) are near the UIT detection limit, and are best considered
as three sigma detections.    The sources BATC 4672 (\bv = 0.46) and BATC 3009 (\bv
= 0.24) have redder optical colors than expected for a hot white dwarf, and
thus might be white dwarf -- red dwarf binaries.

The approximate sensitivity limit of the UIT image of M67 at 1520~\AA\ is 
\mbi\ = 16.6 or $8  \times 10^{-16}$ erg s$^{-1}$ cm$^{-2}$ \AA$^{-1}$.   
Using the field white dwarf luminosity function and scale height (275 pc) of
\markcite{boyle89}Boyle (1989), only 1.26 white dwarfs brighter than this
limit   are predicted to be within the UIT $40'$ diameter field of view at the
Galactic latitude and reddening of M67.     (Since the luminosity function is
given in terms of \mv, we used the \logg\ = 8.0 pure hydrogen model atmospheres
of  \markcite{berg95}Bergeron et al.\ (1995) to compute absolute magnitudes,
\mv, and \mbi\ -- V colors as a function of \teff.)      This low contamination
rate is consistent with a general lack of white dwarf candidates in other
random UIT fields.     In view of this low expected contamination rate, and the
clustering of the white dwarf candidates toward the cluster center, most of the
stars in Table \ref{tbl-4} are likely cluster members.   Fleming et al.\  do
raise a  caveat about the cluster membership of MMJ 5670 and MMJ 5973.     They
compute absolute magnitudes for the two stars using a mass-radius relation, 
and the \teff\ and \logg\ values derived from the best-fit model atmosphere,
and find that the very hot DA star MMJ 5670 is 1 mag too bright, and the DB
star MMJ 5973 is 0.7 mag too faint to be consistent with a cluster distance
modulus of m -- M = 9.7.   However, there are several large uncertainties in
these absolute magnitude determinations.   The S/N of the white dwarf spectra
are low ($\sim 20$), and there are known problems with  model atmospheres for
both DB stars and very hot DA stars.   The optical photometry of MMJ 5973 is
near the detection limit of MMJ93, and there is a 0.4 mag difference between
MMJ93 and \markcite{gill91}Gilliland et al.\ (1991) in the photometry of MMJ
5670.     Nevertheless, the absolute magnitude differences  are still 
uncomfortably large, and proper motion studies of the two stars would be very
desirable to verify cluster membership.  

M67 has a core radius of $5.2'$ and a tidal radius of about $75'$
(\cite{zhao96}), whereas the UIT image has a diameter of $40'$ and is offset by
$4.2'$ from the center of M67.   To estimate the incompleteness of the UIT white
dwarf sample, we note that  85\% of the 222 stars
with membership probabilities greater than 90\% in the wide field
astrometric study of \markcite{zhao93}Zhao et al.\ (1993), are located within UIT field of
view.    Thus, one or two white dwarf
cluster members brighter than the UIT detection limit might be missing from 
Table \ref{tbl-4}. 

How many white dwarfs brighter than the UIT detection limit are expected in
M67, based on the cluster size and age?       Using pure hydrogen  model
atmospheres with \logg\ = 8.0, and the mass-radius relation of 
\markcite{wood95}Wood (1995), the temperature of a 0.6 \msun\ WD at the
distance and reddening of M67 at the UIT detection flux limit is computed to be
21,300 K.    From the white dwarf cooling models of  \markcite{wood95}Wood
(1995), the time for a 0.6 \msun\ WD to cool to this temperature is about 60
Myr.    Thus,  all the white dwarfs in M67 less than 60 Myr old should
detectable on the  UIT image, including those that would be optically hidden in
binary systems with late-type companions.   (Only possible white dwarf
companions of the hottest blue stragglers would remain undetected on the UIT
image.)     The question now becomes -- how many white dwarfs should have been
created in M67 in the last 60 Myrs?   A quick  estimate of about seven white
dwarfs can be made by simply comparing with the seven observed red clump
stars, which  are expected to have a similar lifetime ($\sim 70$ Myr,
\cite{trip93}). Alternatively, \markcite{dorm95}Dorman et al. (1995) estimate
that there should be  0.076 stars leaving the main  sequence per Gyr per
L$_{V}^{\odot}$ for a 4 Gyr solar metallicity isochrone.    Multiplying this
number by 0.060 Gyr and the cluster visible luminosity of L/L$_{V}^{\odot}$ =
1660 (\cite{batt94}) gives an estimate of 7.6 WDs.    Thus, the predicted
number of WDs is in reasonable agreement with the eight detected white dwarf
candidates.    We conclude that the stellar evolution timescale and the hot
white dwarf cooling timescale are not in major disagreement, and that most of
the young ($<$ 60 Myr) white dwarfs have not yet evaporated from the cluster. 
Note that although there is a marked deficit of 0.6 \msun\ {\em main-sequence} 
stars in M67 (MMJ93), no evaporation of the white dwarfs is expected on a
time scale of 60 Myr (\cite{vhipp98}).     Stronger
assertions will be possible once spectra and proper motion studies have been
obtained for all the white dwarf candidates.     

Because the relatively small population of M67 does not include any post-AGB
or hot HB stars, its integrated ultraviolet spectrum will be dominated
by the hot white dwarfs at wavelengths shortward of Lyman $\alpha$.     Would white
dwarfs be detectable in the integrated ultraviolet spectra of galaxies?
\markcite{bica96b}Bica et al.\ (1996a,b) pointed out the presence of 
features at 1400~\AA\ and 1600~\AA\ in IUE spectra of some early-type galaxies,
and suggested that these  might correspond to the Lyman $\alpha$ satellite
lines  observed at these wavelengths in intermediate  temperature (9000 -
18,000 K) DA white dwarfs (\cite{kiel95}).    However, these features are not
present in spectra of the same galaxies observed with the Hopkins Ultraviolet
Telescope  (HUT: \cite{brown97}).   In addition, the satellite lines at 1400
and 1600~\AA\ are not unique to white dwarfs, but  also appear in the spectra
of low-metallicity A-type and horizontal branch  stars (\cite{hol94}). 
Nevertheless, the population synthesis models of  \markcite{magris93}Magris \&
Bruzual (1993) do suggest that white dwarfs are a small but non-negligible
contributor to the ultraviolet spectrum of ellipticals with a small UV
upturn.    They find that, at 1750~\AA, the contribution of white dwarfs is
10\% of the contribution of hot post-AGB stars in galaxies older than 8 Gyr.

\begin{figure} 
\plotone{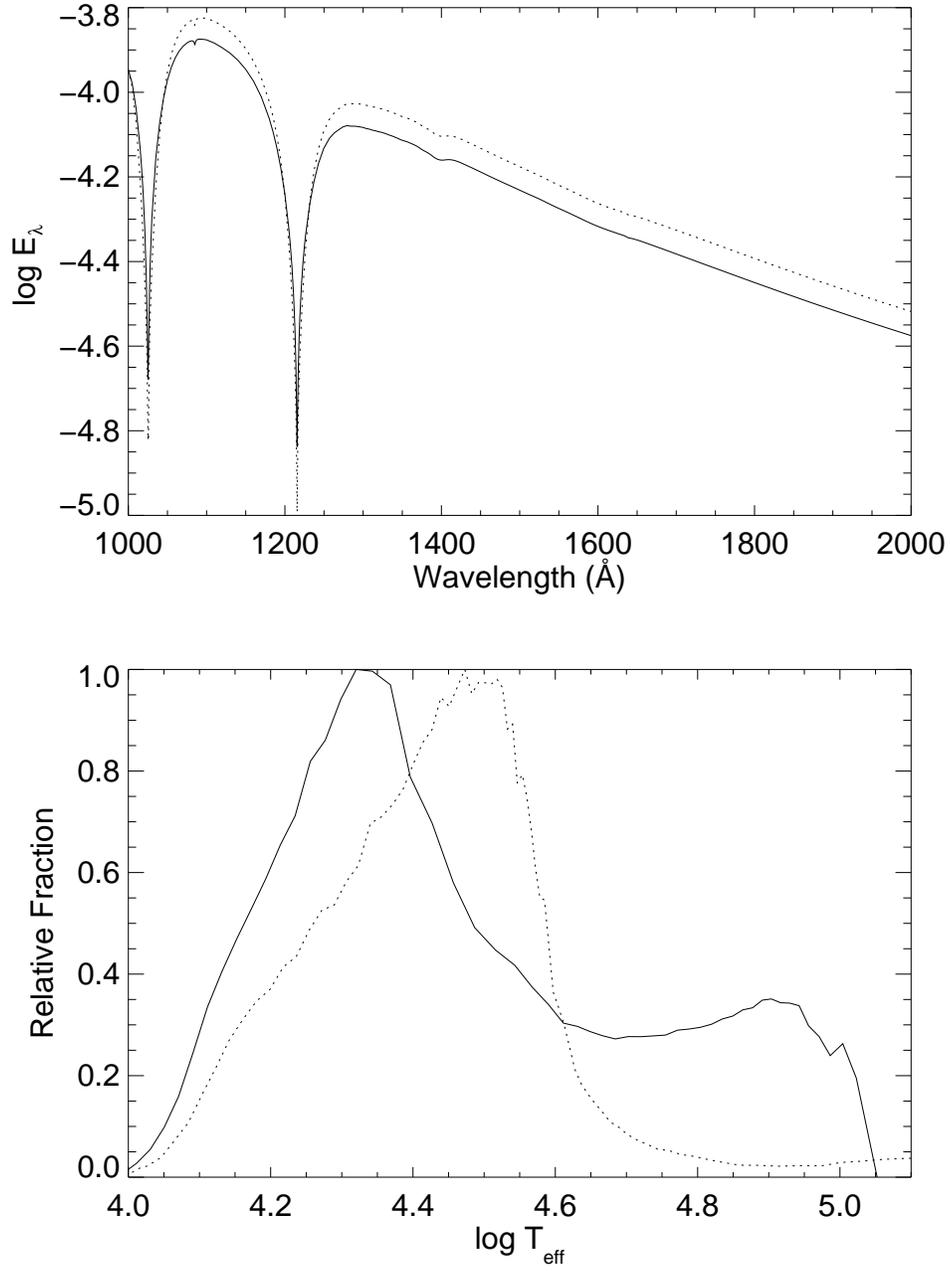}
\figcaption{(a) The ultraviolet spectrum of a
0.6 \msun\ (solid line) and 0.9  \msun\ (dotted line) white dwarf integrated
over its cooling track. (b) the relative contribution per unit $\log$ T to the
integrated white dwarf flux at 1500~\AA\ for cooling white dwarfs of masses 0.6
\msun (solid line) and 0.9 \msun\ (dotted line). \label{fig-5} }  
\end{figure}

We have performed our own estimate of the white dwarf contribution to the
ultraviolet spectrum of an old stellar population.  We use the carbon-core
cooling models of \markcite{wood95}Wood (1995) with thick hydrogen and helium
envelopes, and the pure hydrogen model atmospheres of \markcite{berg95}Bergeron
et al.\ (1995). To allow an easy comparison of the contribution of white dwarfs
with that of other ultraviolet sources, we follow \markcite{dorm95}Dorman et
al.\ (1995) and compute the energy E$_{\lambda}$ integrated over the cooling
curve in units of  L$_{V}^{\odot}$ Gyr \AA$^{-1}$, where L$_{V}^{\odot} = 4.511
\times 10^{32}$ ergs s$^{-1}$ is the visible solar luminosity.   Figure
\ref{fig-5}a shows the resultant ultraviolet spectrum for both a 0.6 \msun\ 
and a 0.9 \msun\ cooling white dwarf.  The time required for the 0.9 \msun\
model to cool to 20,000 K is 190 Myr, whereas only 75 Myr are needed for a 0.6
\msun\ to cool to this temperature.  Despite this large difference in cooling
age, the integrated fluxes,  E$_{\lambda}$, along the two tracks differ by only
$\sim 15$\% outside of the Lyman lines, because the radii in the more massive
model are typically 30\% smaller, which partially compensates for the longer
cooling time.    Other complications, such as the presence of DB white dwarfs,
and variations in the envelope mass, probably do not cause flux variations much
larger than 15\%. Figure \ref{fig-5} also shows that the main contribution to
the integrated spectrum at 1500~\AA\ comes from white dwarfs with \teff\ $>$
20,000 K, so that the Lyman $\alpha$ satellite line at 1400~\AA, characteristic of
cooler DA white dwarfs, is barely detectable.    The integrated energy,  along
the 0.6 \msun\ track at 1500~\AA\ is  $\log$ E$_{1500} = -4.23$ L$_{V}^{\odot}$
Gyr \AA$^{-1}$.     By comparison, \markcite{dorm95}Dorman et al.\ (1995)
estimate $\log$ E$_{1500} = -2.97$ L$_{V}^{\odot}$ Gyr \AA$^{-1}$ for their
lowest core mass (0.546 \msun) post-AGB track, and $\log$ E$_{1500} = -1.6$
L$_{V}^{\odot}$ Gyr \AA$^{-1}$ for typical EHB tracks.    Since the use of a
low core mass probably somewhat overestimates the true post-AGB contribution,
we confirm the result of Magris \& Bruzual (1993) that white dwarfs contribute
a small, but non-negligible ($\sim$ 10\%) fraction of the ultraviolet flux due
to post-AGB stars.     However, the EHB population will dominate the UV
spectrum, even if only a small fraction ($\sim$ 5\%) of core helium-burning
stars follow EHB tracks. 

\section{NGC 6791}

NGC 6791 is noteworthy for being among the oldest, most populous, and most
metal-rich of the Galactic open clusters (\cite{friel95}).   
\markcite{lieb94}Liebert et al.\ (1994) discovered another noteworthy property
of NGC 6791 with their spectroscopic detection in the cluster of an sdO star
and four sdB stars with  24,000 K $<$ \teff\ $<$ 32,000 K.    They speculated
that these  were the metal-rich analogue of the hot HB stars found in globular
clusters, and that the existence of these stars in NGC 6791 was connected with
the super-solar metallicity of the cluster.     Recent stellar evolution models
of high-metallicity, low-mass stars provide support for this idea, and have
also been used to show that such metal-rich hot HB stars could be primarily
responsible for the UV-upturn observed in the spectra of elliptical galaxies
(\cite{bress94}, \cite{dorm95}, \cite{yi97b}).    

Twenty stars are detected on the UIT image of NGC 6791 but twelve of these are
certain non-members, either because they are identified with bright (V $<$13)
field stars, or because they are more than $10'$ from the cluster center.   The
detected probable cluster members include the five hot subdwarfs discussed by
Liebert et al. (B2, B3, B4, B5, and B6), plus the additional two sdB/O
candidate stars, B9 and B10, reported by Kaluzny \& Rucinski (1995).    The
star B1, suspected to be a cooler (\teff\ = 15,000 K) member blue HB star by 
Green et al.\ (1998a), is detected on the UIT image at the three sigma level.   
The other two blue HB stars studied by Green et al., K1943 and K5695, are too
cool to be detected with UIT.    Neither of the cataclysmic variables, B7 and
B8, reported by \markcite{kal97}Kaluzny et al.\ (1997) are detected with
UIT.    Even during their outburst phase, these cataclysmic variables have
redder optical colors than the sdB/O candidates, and probably remain below the
UIT detection limit of \mbv\ $\sim$ 16.1.   Table \ref{tbl-5} presents the UIT photometry
of the possible cluster stars, along with the visible photometry from Kaluzny
\& Rucinski (1995).     

\begin{figure} 
\epsscale{0.85}
\plotone{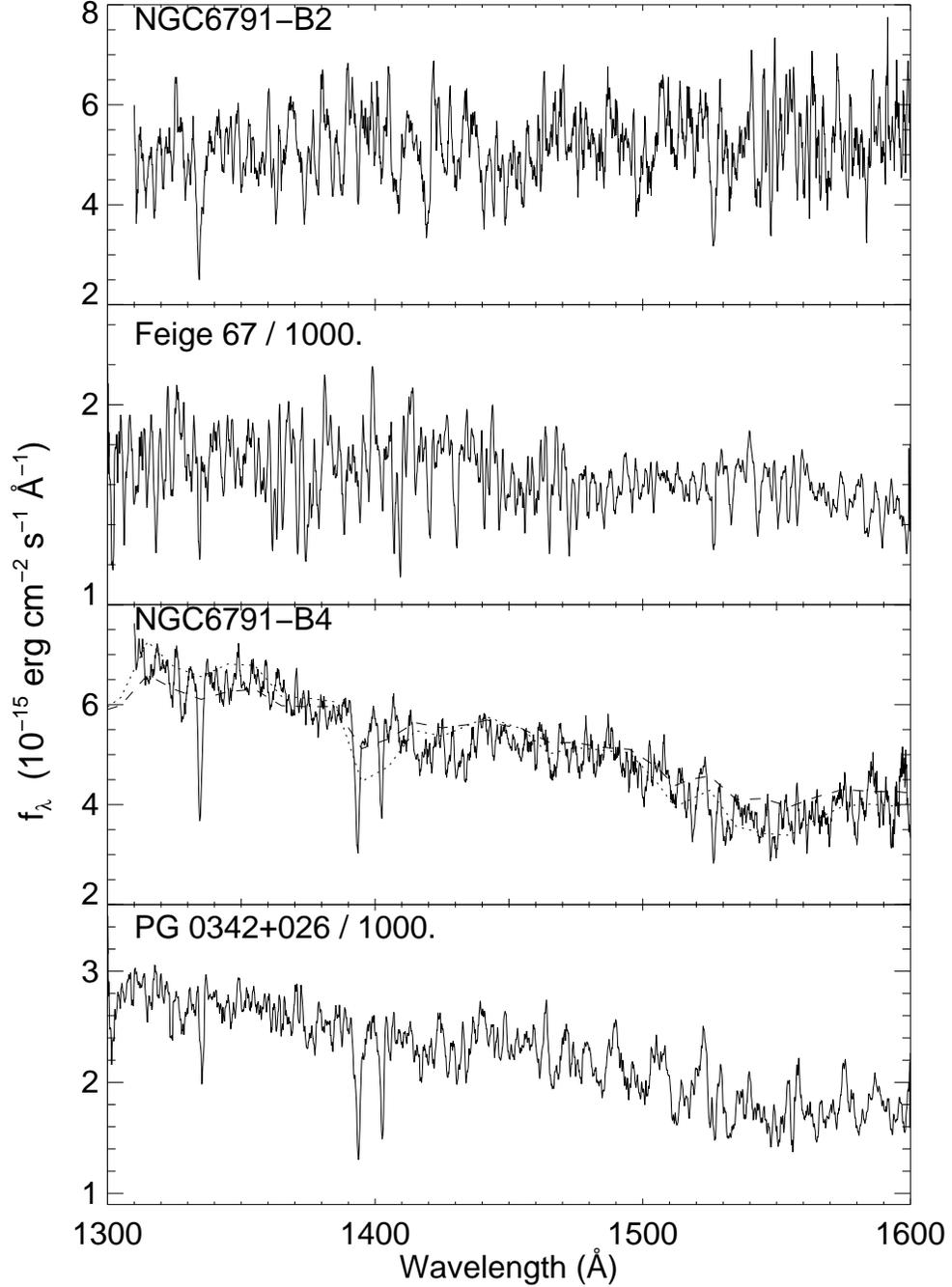}
\figcaption{GHRS low-resolution spectra of NGC
6791-B2 and NGC 6791-B4 are shown along with IUE spectra of the field sdO star
Feige 67 (reddened by E(\bv) = 0.17), and the field sdB star PG 0342+026
(reddened by E(\bv) = 0.06).
The dotted line in the NGC 6791-B4 panel shows a Kurucz
model with \teff\ = 26,800 K, \logg\ = 5.0, and [Fe/H] = 0.3, while the dashed
line shows a model with \teff\ = 25,200 K, \logg\ = 5.0, and [Fe/H] = --0.5.   
The flux scale for the field stars Feige 67 and PG 0342+026 is 
divided by 1000. \label{fig-6}}
\end{figure}

Figure \ref{fig-6} shows archival GHRS low-dispersion (G140L) spectra covering the
wavelength region  1300 -- 1600~\AA\ of the sdO star B2 and the sdB star B4.   
Also shown are high-dispersion \IUE\ spectra of the field sdO star Feige 67
(SWP 20488), and the field sdB star PG 0342+026 (SWP 27466), smoothed to the
approximate GHRS resolution, and reddened to match the NGC 6791 reddening.  
From ultraviolet and optical spectrophotometry, \markcite{thies95}Theissen et
al.\ (1995) found \teff\ = 25,000 K, \logg\ = 5.25 and E(\bv) = 0.10, for PG
0342+026, while Saffer et al.\ (1994) found \teff\ = 26,200 K, \logg\ = 5.63
from fitting the Balmer lines.    The best-fit Kurucz model atmosphere
(normalized to the V magnitude and reddened by E(\bv) = 0.17) to the GHRS 
spectrum of B4 has  \teff\ = 26,800 K, \logg\ = 5.0, and [Fe/H] = 0.2.    

While a good fit to B4 can also be achieved using a solar metallicity model, 
Figure  \ref{fig-6} shows that a low metallicity ([Fe/H] = --0.5) model does
not match the depressed continuum longward of 1500~\AA, which is primarily due
to the lines of \feiii\ and  \feiv\ (\cite{brown96}).\footnote{Note that the Kurucz models
use a solar iron abundance of  -4.37 dex from \markcite{angr89}Anders \&
Grevesse (1989), whereas the current best estimate for the solar iron abundance
is -4.50 dex (\cite{grev96}).}   This result is of some
importance because abundance depletions are known to exist in sdB stars, due to
diffusion and other particle transport processes in the high-gravity
atmosphere.     For example,   \markcite{lieb94}Liebert et al.\ (1994) found
the sdB stars in NGC 6791 (including B4) to be helium-poor, as is typical of
field sdB stars (\cite{saff94}), and \markcite{lam87}Lamontagne et al.\ (1987)
summarizes the evidence for depletions of carbon and silicon in the field sdB
stars (also see Brown et al.\ 1996).    Nevertheless, the carbon and silicon
depletions are  smaller for the sdB stars than for hotter ($>$ 30,000 K) stars,
and other major ultraviolet opacity sources such as iron are not depleted.
Thus, our rough analysis of the GHRS spectrum of B4 suggests that ultraviolet
population synthesis models which incorporate evolutionary tracks of 
metal-rich hot HB  stars should also adopt stellar atmosphere models of an
equal metallicity.   A more detailed comparison of the GHRS spectrum of B4 with 
spectral synthesis models would be desirable to verify this point. 

\markcite{lieb94}Liebert et al.\ (1994) did not derive atmospheric parameters
for the sdO star B2, but showed that its optical spectrum is similar in
appearance to  the hot (\teff\ $\sim$ 60,000 K) field sdO stars Feige 67 and
Feige 110.    Figure \ref{fig-6} shows that Feige 67 and B2 have a similar
continuum slope in the 1300 -- 1600~\AA\ region, and many absorption lines are
in common, although the spectra clearly differ in many details.   Due to the
large number of lines of ionized iron and nickel in the  ultraviolet spectra of
sdO stars, a quantitative analysis requires fully line-blanketed NLTE model
atmospheres, which have only recently become available (e.g.\ \cite{lanz97},
\cite{haas96}).   Becker \& Butler (1995) derived \teff\ = 70,000 K, \logg\ =
5.2 for Feige 67, and found it to have a low surface abundance of helium
(N(He/H) = 0.05), but a high iron abundance ([Fe/H] = 1.0).    We do not
attempt to further estimate the atmospheric parameters of B2, but limit
ourselves to the following remarks.  First, the identification of B2 as an
AGB-manqu\'e star may be suspect, because the evolutionary tracks of
AGB-manqu\'e stars do not spend much time at such high effective temperatures
($\sim$ 60,000~K). Second, despite the higher \teff\ of the sdO stars in Figure
\ref{fig-6}, they have flatter continua than the two sdB stars, mostly likely
due to the large number of  \fev\ lines in the  1300 -- 1600~\AA\ region
(\cite{beck95}).    Thus, accurate population synthesis of the UV-upturn in
elliptical galaxies may need to include NLTE line-blanketed model atmospheres
to properly model AGB-manqu\'e stars.    Unfortunately, the few field sdO stars
which have had a detailed model atmosphere analysis of their ultraviolet
spectra have shown a wide variety of strong elemental (including iron) under-
and over-abundances (\cite{lanz97}, \cite{haas96}, \cite{beck95}), most likely
due to the effects of radiative levitation and diffusion.    The question of
the appropriate model atmosphere to use for population synthesis of the
ultraviolet spectra of metal-rich AGB-manqu\'e stars remains a problematic
one.        

The UIT photometry is especially useful for estimating the luminosities of the
sdB stars, since the bolometric correction at 1620~\AA\ for sdB stars is
relatively small.    
The effective temperature  and luminosity
of the sdB stars B3, B4, B5, B6, and B9 as derived from the UIT photometry is
tabulated in columns 5 and 6 of Table \ref{tbl-5}. 
(No spectra exist for the star B9, but its ultraviolet and visible magnitudes
are similar to those of the four spectroscopically identified sdB stars.)
Temperatures are estimated by finding the Kurucz model with [Fe/H] = 0.2, and
\logg\ = 5.0, which best matches the observed \mbv\ -- V color, for an assumed
reddening of E(\bv) = 0.17.        Luminosities are then derived from the \mbv\
magnitude, using the (small) bolometric correction for the adopted Kurucz model.
The values of \teff\ derived from \mbv\ -- V are consistently about 1200 K 
hotter than those derived by Liebert et al.\ (1994) using (pure hydrogen) model
fits to the Balmer lines.

Table \ref{tbl-5} shows that the five sdB stars in NGC 6791 have a narrow
luminosity  range, $1.10 < \log$L/\lsun\ $< 1.27$.    This luminosity range is 
consistent with that predicted for metal-rich hot HB stars  by 
\markcite{dorm93}Dorman et al.\ (1993) and \markcite{yi97a}Yi et al.\
(1997a).   For example, the hottest models of Yi et al.\ (with Z = 0.04 and a
main-sequence helium abundance Y = 0.31), have a luminosity range during core
helium burning of $1.05 < \log$L/\lsun\ $<$ 1.28.   The observed \teff\ range
of the sdB stars is perhaps slightly higher than predicted, since the \teff\ of
the hottest HB models decreases slightly with increasing metallicity
(\cite{dorm93}).    The hottest models of Yi et al.\ and Dorman et al.\
(with envelope masses less than 0.005 \msun) with Z = 0.04 have \teff $\sim$
23,000~K.

The narrow range of luminosities of the sdB stars in NGC 6791 rules outs
certain binary star mechanisms for their origin, such as the merger of helium
white dwarfs (\cite{iben90}), since these predict a wide range of helium core
masses, and hence luminosities.   However, \markcite{green98a}Green et al.\
(1998a) found that B1 and the other two blue HB stars in NGC 6791 are
spectroscopic binaries.      (The sdB stars are too faint for a radial velocity
study, so their binary status is unknown.)    The situation is reminiscent of
that of the field sdB stars, which occupy a narrow region in the \teff\ --
\logg\ diagram,  consistent with single star evolution (\cite{saff94}), but for
which there is increasing evidence for a large binary population
(\cite{allard94}, \cite{theis95}, \cite{green98b}). One binary star mechanism
consistent with a narrow range of gravity or luminosity is the scenario
discussed by \markcite{meng76}Mengel et al.\ (1976, also see \cite{bailyn92}),
in which an evolving red giant loses mass to a companion, so that it arrives on
the horizontal branch with a small envelope mass.  In effect, the mass transfer
to the companion substitutes for the high mass-loss rate required for the
production of sdB stars in single star evolution models.    One difficulty with
both the high mass-loss rate and mass-transfer scenarios is the amount of
fine-tuning required (\cite{lieb94}), since a red giant which loses too much
mass never reaches the helium flash, and instead ends up as a helium white
dwarf (but see \cite{dcruz96}).    Note that the maximum envelope mass
for which a  star still follows AGB-manqu\'e tracks increases
with increasing metallicity  (\cite{yi97a}, \cite{dorm95}), so that the
binary star mechanism for the production of sdB stars might still show a
metallicity dependence.      

The brightest source in NGC 6791 at 1620~\AA\ is B10, which has \mbv\ = 13.60. 
\markcite{karu95}Kaluzny \& Rucinski (1995) report that B10 is a blend of two stars with 
$\Delta$V $\sim$ 2.0 mag and an angular separation of about 0.7\arcsec.   The
photometry in Table \ref{tbl-5} is from the 2.1-m telescope observations of Kaluzny \&
Rucinski (1995), in which the photometry of the hot star is better isolated.
Nevertheless, the hot star might still have a composite spectrum because its 
ultraviolet color (\mbv\ -- V = --2.68) is somewhat inconsistent with its
relatively red \bv\ color (\bv\ = $0.014 \pm 0.014$).   
\markcite{green98a}Green et al.\ (1998a) obtained three spectra of B10 in which
the neighboring cool star apparently dominates the spectrum.   They find a
radial velocity consistent with cluster membership, and no evidence for
velocity variations, although it is not clear whether these results also apply
to the hot star.    Minimum values of \teff\ $>$ 22,900 K and $\log$L/\lsun\ $> 1.75$
for B10 can be derived from the UIT photometry by neglecting any contribution
to the V light from a possible red companion.   B10 appears to be a good
candidate for a AGB-manqu\'e star.

\section{NGC 188}

Seven stars are detected on the UIT image of the old ($\sim 6$ Gyr) 
solar-metallicity open cluster NGC 188, but  four of these are certain
non-members (Table \ref{tbl-6}). The early photometry of
\markcite{sand62}Sandage (1962) included the hot star II-91 (V = 13.82,  \bv =
--0.22), but only recently has this star been shown to be a probable
proper-motion cluster  member (\cite{din96}), and to have an sdB spectrum
(\cite{green98a}).    Green et al.\ also report that II-91 is a spectroscopic
binary.   As with the sdB stars in NGC 6791, the effective temperature 
and luminosity of II-91 can be estimated from its UIT magnitude and \mbi\ -- V
or \mbv\ -- V color, for the adopted cluster distance (1700 pc) and  reddening
(E(\bv) = 0.12).    Use of either the \mbv\ or \mbi\ magnitude yields a
temperature of about 30,000 K and luminosity of $\log$ L/\lsun\ = 1.14.   
Thus, the temperature and luminosity of II-91 falls into the same range found
for the sdB stars in NGC 6791.  

The brightest source in NGC 188 at 1520~\AA\ is the star D702, from  the
catalog of \markcite{din96} Dinescu et al.\ (1996), who give V = 13.70 and \bv\
= 0.26 from their photographic photometry, and also assign a 80\% probability
of cluster membership on the basis of its  proper motion.   Dinescu et al.\ also
remark, however, that D702 is part of a blended image, so that it has large
proper motion errors, and possibly erroneous photometry.   D702 is also E43
in the catalog of \markcite{cap90}Caputo et al.\ (1990), who give V = 14.04 and
\bv\ = 0.28 from their CCD photometry.    Here we adopt the photometry (V =
14.18, \bv\ = 0.26) quoted by \markcite{green98a}Green et al.\ (1998a).   The ultraviolet
photometry  shows that D702 must have a composite spectrum, because its
relatively red \bv\ color is inconsistent with its blue ultraviolet colors
(\mbi\ -- V = --1.89; \mbi\ -- \mbv\ = --0.11).     A minimum value of \teff\ =
15,000 K can be derived for D702 from the observed  \mbi\ -- V color by
neglecting any contribution of a red companion to the V magnitude; the \teff\
could be much higher, if the V magnitude includes significant contamination
from the companion. One possible decomposition of D702 consistent with the
visible and ultraviolet photometry is that it consists of a turnoff star with V
= 14.9 and \bv\ = 0.69, and a hot companion with V = 14.9 and \bv\ = --0.04
and  \teff\ = 15,000 K.     Alternatively, the hot companion could be similar
to II-91 with V = 16.2 and \teff\ = 30,000 K, in which case the primary would
be a blue straggler with V = 14.36 and \bv\ = 0.36.    

Unlike the case for M67, none of the blue stragglers in NGC 188 or NGC 6791 are
detected with UIT (with the possible exception of D702), nor would any
detections be expected.   This is partly because these clusters are more
distant and more reddened, and their UIT exposures shallower than that of M67,
but also because these older clusters have a blue straggler population with
redder \bv\ colors. 

A final UIT source, located $13'$ from the cluster center, and designated
here as UIT-1, could not be identified with any sources in the NGC 188 star
catalogs. However, it is located $1.4''$ from the source U1725\_00042710 in
the  USNO-SA1.0 sky survey of \markcite{monet96}Monet et al.\ (1996).     In their
photometric system defined by the IIIa-J and IIIa-F emulsions, U1725\_00042710
has the very blue colors, r=18.7 and b = 18.0.   However, a 4000 -- 5000~\AA\
spectrum of this star obtained at Calar Alto in 1996 August (S. Moehler,
personal communication) showed a very red continuum with no evidence of broad
Balmer lines.   Whether this discrepancy indicates an eclipsing binary, or is
due to a misidentification is currently unknown.    The star is robustly
detected in both UIT filters, and the ultraviolet color \mbi\ -- \mbv\ = --0.17
indicates a hot ($>$ 20,000 K) spectrum.

The CMD of NGC 188 shown by  \markcite{din95}Dinescu et al.\ (1995) has five
red clump stars, so the two sdB candidates detected by UIT provide 28\% of the
HB population.  The CMD of NGC 6791 given by  \markcite{kal95}Kaluzny \&
Rucinski (1995) shows about 23  red clump stars, so the eight hot stars
detected by UIT, plus the two cooler blue HB stars discussed by
\markcite{green98a}Green et al.\ (1998a), form about 30\% of the HB
population.   By comparison, \markcite{dorm95}Dorman et al. (1995) found that
they could fit the ultraviolet spectra of even the strongest UV-upturn galaxies
if only 25\% of the core helium-burning stars followed EHB tracks, while 
\markcite{brown97}Brown et al.\ (1997) needed an EHB fraction of at most 10\%
to model the HUT spectra of six ellipticals.    Thus, the integrated spectra of
NGC 6791 and NGC 188 would likely show a strong UV-upturn, although the 
present incompleteness of the cluster membership information prevents us from
computing  semi-empirical integrated spectra for the two clusters, as we did
for M67 in Section 3.1.

The similarity of the EHB fraction for NGC 188 and NGC 6791  might suggest
that the super-solar metallicity of NGC 6791 is not an essential ingredient in the
origin of its hot subdwarf population.    However, there are several reasons
why this conclusion might be premature.    First, high percentage of sdB stars
in NGC 188 could simply be an artifact of small number statistics. Second,
apart from NGC 188,  no sdB stars have yet been found in any other open cluster
with a metallicity less than or equal to solar.     Finally, both sdB
candidates in NGC 188 are in binary systems, and, as pointed out by Green et
al.\ (1998a), NGC 188 is known to have  a rich binary population,  whereas NGC
6791 appears to have a relatively low binary fraction.        

\section{Summary}

There are at least three ways in which Galactic open clusters are an inadequate
model for  understanding the stellar populations of old galaxies.    First, the
open clusters have experienced dynamical interactions which can alter the
stellar population mix; for example, by leading to evaporation of low-mass
stars while concentrating the blue stragglers toward the cluster center.   
Second, the higher mean stellar density of open clusters might result in  
enhanced stellar interactions and possibly increased blue straggler or hot
subdwarf formation.
Finally, due to the relatively small population of  open clusters, the rapid
evolutionary phases may not be adequately sampled.  For example, hot post-AGB
stars are believed to be a significant (although not the dominant)  contributor
to the observed UV-upturn in elliptical galaxies, but consistent with their
short ($\sim 10^5$ yr) lifetime, there are no post-AGB stars present in M67,
NGC 188, or NGC 6791.    And while hot subdwarfs have now been detected in NGC
188 and NGC 6791, their total number ($<8$) remains uncomfortably low to draw
any robust conclusions. 

Keeping this limitations in mind, we summarize below
the main results derived from the UIT observations of M67, NGC 188, and NGC
6791, and their implications for the study of the integrated ultraviolet
spectra of old stellar populations.

\begin{enumerate}

\item  The UIT image of M67 is dominated by the eleven detected blue
stragglers; in particular, the blue straggler F81 contributes 90\% of the
integrated flux of M67 at 1520~\AA.    The ultraviolet flux of the two blue
stragglers,  F90 and F131, is significantly higher than that predicted on the
basis of optical photometry, and probably indicates the presence of hot,
subluminous companions.   A semi-empirical calculation of the integrated
ultraviolet spectrum of M67 shows that,  even when neglecting the possibly
anomalous star F81, the blue stragglers dominate at wavelengths shorter than
$\sim$ 2600~\AA.   As pointed out by Spinrad et al.\ (1997), neglect of blue
stragglers in population synthesis model fits of the rest-frame near-ultraviolet
spectra of a galaxy will result in an underestimate of
the galaxy age.

\item Eight white dwarf candidates are identified in the UIT image of M67,
including the core-helium white dwarf companion of the yellow giant S1040.   
Optical spectroscopy of two of these sources has been obtained by
\markcite{flem97}Fleming et al.\ (1997): G152 is a hot (\teff $\sim$ 68,000 K)
DA white dwarf, and MMJ 5973 is cooler (\teff\ $\sim$ 18,000 K) DB white dwarf.
The number of white dwarf candidates is in reasonable agreement with that
expected from theoretical white dwarf cooling models and a cluster age of 4
Gyr.    The integrated ultraviolet flux at 1500~\AA\ along a white dwarf
cooling model is $\log$ E$_{1500} = -4.23$ L$_{V}^{\odot}$ Gyr \AA$^{-1}$, and
the contribution of white dwarfs to the integrated spectra of old galaxies is
roughly 10\% of that expected from hot post-AGB stars.

\item Eight probable cluster members are detected on the UIT image of NGC 6791,
including the five hot subdwarfs studied spectroscopically by Liebert et al.\
(1994), and the two additional sdB/O candidates, B9 and B10, reported by
\markcite{kal95}Kaluzny \& Rucinski (1995).    Three probable cluster members
are detected on the UIT image of NGC 188, including the sdB spectroscopic
binary II-91.   The star D702 in NGC 188 is probably a composite including a
hot subdwarf, since it has a hot ultraviolet color (\mbi\ -- \mbv\ = --0.11),
but a relatively cool optical color (\bv\ = 0.26).  The derived  luminosity
range, $1.10 < \log$ L/\lsun\ $< 1.27$, of the five sdB stars in NGC 6791, and
II-91 in NGC 188, is consistent with that expected for metal-rich, hot HB
stars.    The fraction of hot HB stars in both clusters is about 30\%,  
implying that the integrated spectra of both clusters should show a pronounced
UV-upturn, as strong as that observed in any elliptical galaxy.

\end{enumerate}

\acknowledgments

We thank the many people involved with the  {\em Astro-2} mission who made
these observations possible.  We thank P. Bergeron for his white dwarf
atmosphere models, A. Milone for providing the latest status of the M67
radial velocity monitoring, and D. Dinescu for providing an electronic version
of her NGC 188 catalog.

\clearpage

\begin{figure}
\figcaption{ A 19\arcmin square section of the UIT image (left) of M67, along
with a matching visible image (right)  from the STScI Digitized Sky Survey.  
The blue stragglers are marked by squares, the white dwarf candidates by
circles, and the yellow giant -- white dwarf binary S1040 is marked with a
triangle. The stellar centroid of M67 is marked with a plus sign on the UIT
image. \label{fig-1}}  
\end{figure}

\clearpage

\clearpage

\begin{deluxetable}{rrrlrl} 
\tablenum{1}
\tablecaption{Observed Clusters \label{tbl-1}}
\tablehead{
 \colhead{Name}  & \colhead{Age (Gyr)} & \colhead{d (pc)} &  \colhead{E(\bv)}
  & \colhead{[Fe/H]} & \colhead{Ref}  }

\startdata
 M67      &  4     &   820 & 0.025  &  --0.1 & Carraro et al.\ (1996)  \nl
 NGC 188  &  6 & 1680 & 0.12   &  0.0 &   Dinescu et al.\ (1995) \nl
 NGC 6791 &  7                 & 4200 & 0.17  &   +0.5 &  Kaluzny \& Rucinski (1995) \nl
\enddata
\end{deluxetable}

\begin{deluxetable}{rrrrr}
\tablenum{2}  
\tablecaption{UIT Observing Log. \label{tbl-2}}
\tablehead{
 \colhead{Name}  & \colhead{Filter} & \colhead{Time (s)} &  
\colhead{RA (2000)}   & \colhead{Dec (2000)}  }
\startdata

 M67      &  B1 &  1210 & 08 51 11.0 & +11 47 14        \nl
 NGC 188  &  B1 &   632 & 00 48 41.6 & +85 20 58        \nl
          &  B5 &   316 &            &                  \nl
 NGC 6791 &  B5 &   984 & 19 20 55.0 & +37 39 59        \nl
\enddata

\end{deluxetable}

\begin{deluxetable}{rrrrlrrcr}
\tablenum{3}  
\tablecaption{M67 Blue Stragglers  \label{tbl-3}}
\tablehead{ \colhead{ID} & \colhead{Sanders} & \colhead{V} & \colhead{\bv} &  
\colhead{\mbi} & \colhead{\teff} & \colhead{\logg} &\colhead{\mbi (pred)} &
 \colhead{Comment} } 

\startdata

81  &  977 & 10.03 & --0.09  & \nodata      & 12750  & 4.26 &  8.17   &       \nl
280 & 1434 & 10.70 & 0.10    & 12.06 &  8600  &  3.89 &12.25  &         \nl
156 & 1066 & 10.99 & 0.11    & 12.10 &  8680  &  4.13 &12.39  &         \nl
153 & 968  & 11.28 & 0.13    & 13.15 &  8490  &  4.31 &13.04  &   Am    \nl
55  & 752  & 11.32 & 0.17    & 15.61 &  7620  &  4.00 & 14.90 &   P = 1003d    \nl
238 & 1267 & 10.94 & 0.22    & 14.20 &  8010  &  3.85 & 13.72 &   P = 846d      \nl
190 & 1284 & 10.94 & 0.25    & 14.47 &  7750  &  3.78 & 14.16 &   P = 4.2d, $\delta$ Sct \nl
185 & 1263 & 11.06 & 0.26    & 14.31 &  8290  &  4.14 & 13.18 &            \nl
184 & 1280 & 12.26 & 0.29    & 15.28 &  8090  &  4.12 &15.33  &  $\delta$ Sct \nl
261 & 1466 & 10.60 & 0.34    & 14.32 &  \nodata  &  \nodata  & \nodata  &   Nonmember? \nl
131 & 1082 & 11.25 & 0.41    & 15.16 &  6930  &  4.09  & 16.47  &           \nl 
90  &  975 & 11.08 & 0.43    & 16.18: & 6490  &  3.44   & 18.04  & P = 975d   \nl
\enddata

\end{deluxetable}

\begin{deluxetable}{lcrrrl}
\tablenum{4}  
\tablecaption{M67 White Dwarf Candidates  \label{tbl-4}}
\tablehead{ \colhead{ID} & \colhead{V} & \colhead{\bv} &  
\colhead{\mbi} & \colhead{d($'$)} &\colhead{Notes}  }

\startdata

MMJ 5670, G152  & 18.61  & --0.11  & 13.92  & 1.7    & DA1         \nl
MMJ 5973        & 19.66  & --0.10  &  15.61 &  6.7   & DB          \nl
S1040          & 11.52  & 0.82    & 15.82  & 1.0     & G8III + WD  \nl
BATC 4672      & 17.66  & 0.46    & 15.84  & 10.1    &             \nl
BATC 3009      & 17.02  & 0.24   & 16.09   &   14.2  &             \nl
MMJ 6061        & 20.53  & --0.14  & 16.50  &  4.3    &             \nl 
BATC 2776      &   ...     & ...     & 16.58   & 7.5  & $m_{3890} = 19.00$\nl
BATC 3337      & 18.44  & 0.09   & 16.66   &  12.0      &          \nl
\enddata

\end{deluxetable}

\begin{deluxetable}{lrrlrrrl}
\tablenum{5}  
\tablecaption{UIT Photometry in NGC 6791  \label{tbl-5}}

\tablehead{ \colhead{} & \colhead{} & \colhead{} & 
\colhead{} & \multicolumn{2}{c}{UV}  & 
 \multicolumn{2}{c}{Optical\tablenotemark{a}} \\
\colhead{ID}   & \colhead{V}          & \colhead{\bv} &  
\colhead{\mbv} & \colhead{\teff\ (K)} & \colhead{$\log$ L/L$_{\odot}$} & 
\colhead{\teff\ (K)}  & \colhead{Sp.T} } 

\startdata

B1  & 16.97  & 0.07    & 16.20: & \nodata & \nodata     & 15,800 & BHB?  \nl
B2  & 17.43  & --0.12    & 14.47  & \nodata  & \nodata & \nodata &sdO           \nl
B3  & 17.77  & --0.12    &  14.96 &  24,170   & 1.21  & 22,900 & sdB   \nl 
B4 & 17.87 & --0.13    & 14.85  &  26,790  &   1.27 &  25,200 & sdB   \nl
B5 & 17.90 & --0.10  & 15.21     & 22,770  &   1.10 &   21,800 & sdB   \nl
B6 & 17.97 & --0.10    & 15.07   & 25,290   &  1.17 &   23,800 & sdB   \nl
B9 & 18.18  & --0.15 & 14.91 & 30,050   &  1.25    &  \nodata & \nodata \nl
B10 & 16.28 &  0.01 & 13.60 & $>22,660$ & $>1.74$  &  \nodata & \nodata  \nl
\tablenotetext{a}{Spectral type and \teff\ from Liebert et al.\ (1994) or
Green \& Liebert (1998a)}

\enddata
\end{deluxetable}

\begin{deluxetable}{lrrrrl}
\tablenum{6}  
\tablecaption{UIT Sources in the NGC 188 field  \label{tbl-6}}
\tablehead{ \colhead{ID} & \colhead{V} & \colhead{\bv} &  
\colhead{\mbi} & \colhead{\mbv} & \colhead{Comment} } 

\startdata

HD 4041  & 8.87  & 0.17    &   9.96 & ...     &  A2V, nonmember       \nl
HD 4816  & 8.26  & 0.20    &  12.45 & 11.65    &  A5V, nonmember      \nl
Hipp 3354\tablenotemark{a}  & 9.58  & 0.50    &  16.13 &  ...    &   F3/F5V, nonmember       \nl 
Tyc 4619\tablenotemark{a} & 11.28 & 0.08    & 15.13 &  14.38  &    nonmember          \nl
II-91    & 16.31 & --0.17  & 12.55 &  12.78  &  sdB    \nl
D-702    & 14.18 & 0.26    & 12.29 &  12.40   & Dinescu et al.\ (1996)    \nl
UIT-1\tablenotemark{b} &  ...  & ...     & 13.95 &  14.12  &  
USNO 1725-00042710?       \nl
\tablenotetext{a}{Identification and BV photometry from the Hipparcos or Tycho
catalogs (\cite{esa97})}
\tablenotetext{b}{located at 00$^h$ 53$^m$ 0\fs7 +85\arcdeg 25\arcmin 34\arcsec}

\enddata

\end{deluxetable}

\end{document}